# A kinetic model of continuous radiation damage to populations of cells: Comparison to the LQ model and application to molecular radiotherapy

Sara Neira[1,+], Araceli Gago-Arias[1,2,+], Jacobo Guiu-Souto[3], Juan Pardo-Montero[1,2,*]

1. Group of Medical Physics and Biomathematics, Instituto de Investigación Sanitaria de Santiago (IDIS), Santiago de Compostela, Spain.

2. Department of Medical Physics, Complexo Hospitalario Universitario de Santiago de Compostela, Spain.

3. Department of Medical Physics, Centro Oncolóxico de Galicia, A Coruña, Spain.

+ Equal contribution.

**Corresponding author:** Juan Pardo-Montero, Instituto de Investigación Sanitaria de Santiago (IDIS), Servizo de Radiofísica e Protección Radiolóxica, Hospital Clínico Universitario de Santiago, Trav. Choupana s/n, 15706, Santiago de Compostela; Phone: +34 981950969; E-mail: juan.pardo.montero@sergas.es

**Running title:** Kinetic model of continuous radiation damage to populations of cells

**Abstract**

The linear-quadratic (LQ) model to describe the survival of irradiated cells may be the most frequently used biomathematical model in radiotherapy. There has been an intense debate on the mechanistic origin of the LQ model. An interesting approach is that of obtaining LQ-like behavior from kinetic models, systems of differential equations that model the induction and repair of damage. Development of such kinetic models is particularly interesting for application to *continuous dose rate* therapies, such as molecular radiotherapy or brachytherapy. In this work, we present a simple kinetic model that describes the kinetics of populations of tumor cells, rather than lethal/sub-lethal lesions, which may be especially useful for application to *continuous dose rate* therapies, as in molecular radiotherapy. The multi-compartment model consists of a set of three differential equations. The model incorporates in an easy way different cross-interacting compartments of cells forming a tumor, and may be of especial interest for studying dynamics of treated tumors. In the *fast dose delivery limit*, the model can be analytically solved, obtaining a simple closed-form expression. Fitting of several surviving curves with both this solution and the LQ model shows that they produce similar fits, despite being functionally different. We have also investigated the operation of the model in the *continuous dose rate* scenario, firstly by fitting pre-clinical data of tumor response to $^{131}$I-CLR1404 therapy, and secondly by showing how damage repair and proliferation rates can cause a treatment to achieve control or not. Kinetic models like the one presented in this work may be of special interest when modeling response to molecular radiotherapy.

# 1. Introduction

The linear-quadratic (LQ) model to describe the survival of irradiated cells may be the most important and most employed biomathematical model in radiotherapy. According to the LQ model (Lea and Catcheside 1942, Kelleren and Rossi 1972, Fowler 1989), the surviving fraction of a population of irradiated cells depends on the absorbed dose $D$ as:

$$SF(D) = \exp(-\alpha D - \beta D^2) \qquad (1)$$

The linear parameter, $\alpha$, is supposed to model death by lethal damage, and the quadratic parameter, $\beta$, death by the accumulation of sub-lethal damage. While the concepts of lethal and sub-lethal lesions may not have a mechanistic biological interpretation, they provide a mathematically good description of surviving fraction curves. Several modifications of the LQ model have been proposed over the years, aiming at including hypoxia radio-resistance (Wouters and Brown 1997), low dose hypersensitivity (Joiner et al 2001), or high dose modifications (Guerrero and Li 2004, Guerrero and Carlone 2010, Wang et al 2010). There has been an intense debate on the mechanistic origin of the LQ model (see for example McMahon 2018, Zaider 1998, Sachs and Brenner 1998, McMahon and Prise 2019, or Bodgi et al 2016). The most recent hypothesis links the mechanistic origin of the LQ-model to the involvement of the ATM protein in radiation repair (Bodgi and Foray 2016).

Equation (1) explicitly assumes that the administration of dose is fast compared to the rates of damage repair. If this is not the case, and the administration of the dose takes long, the accumulation of sub-lethal damage will cause less effect due to damage repair. The above equation can be written as (O'Rourke et al 2009):

$$SF(D, r(t)) = \exp(-\alpha D - \beta G(r(t), \{\mu\}) D^2) \qquad (2)$$

The factor $G$ is usually called the protraction factor and accounts for damage repair. It depends on the dose rate $r(t)$, and on the set of parameters modelling damage repair, $\{\mu\}$. In the *instantaneous delivery limit* $G$ tends to 1, while in the very low dose rate, $G$ will go to 0. The LQ-model with protraction of dose has been used for example to model the effect of molecular radiotherapy (Dale and Carabe-Fernández 2005, Kalogianni et al 2007, Denis-Bacelar et al 2017).

An interesting approach is that of trying to obtain LQ-like behavior from kinetic modelling, i.e. systems of differential equations that model the induction and repair of damage, including lethal and sub-lethal lesions. This approach, originally developed by Lea and collaborators (Lea 1955, Lea and Catcheside 1942), lead to the development of several classic models (Chadwick and Leenhouts 1973, Tobias 1985, Goodhead 1985, Curtis 1986), reviewed and evaluated by Bodgi et al (2016a) and Sontag (1990). While these models are not mechanistic *per se*, they can provide very useful information on the particular kinetics of radiation death. In addition, if these models are simple enough, closed-form solutions can be obtained and compared to what is expected from the LQ. Radiation damage, repair, and induced cell death are still very active areas of research, including Monte Carlo microdosimetry and generation of DNA damage (Carlson 2008), and new models of damage repair kinetics (Belov et al 2014, McMahon et al 2016, Meylan et al 2017)

In this work, we present a kinetic model along with the principles introduced above. Our model directly describes the kinetics of populations of (tumor) cells, rather than lethal and sub-lethal damage like other models do. This model is similar to that presented in Scheidegger et al (2011). In that paper, the authors introduced *ad hoc* a kinetic model that has the LQ-model as a solution. In our model, we specifically include different cell populations (compartments), and transfer of cells between compartments according to different kinetic processes, which may make it more suitable to describe the evolution of a tumor. We present and describe the model, obtaining a closed-form solution in the *fast delivery limit* that we compare with the LQ model. We use the model to fit many experimental data of surviving fraction versus dose.

We also investigate the behaviour of the model in the *continuous dose rate* scenario, as in molecular radiotherapy (targeted radionuclide therapy). In this type of therapy, a radio-labeled drug with an affinity for the tumor is administered to the patient. The drug is preferentially taken up by the tumor, and radionuclide decay and biological elimination lead to a continuous tumor irradiation with degrasing dose . In such therapies, where damage repair is critical (Dale and Carabe-Fernández 2005, Dale 1985) and kinetic models may find ample application (Barbolosi et al 2017), this work may be of special interest. We use our model to fit experimental data of evolution of tumor volumes after molecular radiotherapy, and we also present a study on the evolution of populations of tumor cells during molecular radiotherapy and tumor control probabilities.

## 2. Methods and Materials

### 2.1. Overview of the model

Let us assume a population of tumor cells, *N(t)*. If there is no treatment, cells will proliferate freely, increasing the population of cells. There is evidence for exponential proliferation slowing down for large tumors, therefore the need to include some sort of *saturation* mechanism in growth models (Kim and Tannock 2005). We consider a logistic-like proliferation term with an exponential rate λ, moderated when the number of cells approaches a given maximum number of cells, $N_{sat}$. Let us assume a radiation treatment is applied, described by a dose rate *r(t)*. We assume that delivered dose can damage cells either lethally or sub-lethally, with a rate proportional to parameters *a* and *b*, respectively, and to the dose rate. This will create two new compartments, doomed cells $N_d$ that have been lethally damaged and will eventually die, and sub-lethally damaged cells $N_s$ that can eventually recover due to damage repair, or can become doomed if they are lethally damaged or accumulate more sub-lethal damage while in that state. We assume that the repair rate of sub-lethally damaged cells is proportional to the number of such cells (therefore exponential). The parameter describing repair is named *μ*. Generally, we can assume that $N_d$ and $N_s$ cells still carry some proliferative capacity, and can proliferate (it is known that doomed cells may still go through some division cycles before dying, the so-called abortive divisions, Dörr 1997), with rate parameters $λ_d$ and $λ_s$ that may be different from that of undamaged cells. Finally, we assume that doomed cells die according to an exponential function with rate *γ*, like in Gago-Arias et al (2016).

### 2.2. The model

The model as described in section 2.1 has the following functional form:

$$\frac{dN(t)}{dt} = \lambda N(t)\left[1 - N_T(t)/N_{sat}\right] + \mu N_s(t) - ar(t)N(t) - br(t)N(t) \tag{3}$$

$$\frac{dN_d(t)}{dt} = \lambda_d N_d(t)\left[1 - N_T(t)/N_{sat}\right] + ar(t)N(t) + ar(t)N_s(t) + br(t)N_s(t) - \gamma N_d(t) \quad (4)$$

$$\frac{dN_s(t)}{dt} = \lambda_s N_s(t)\left[1 - N_T(t)/N_{sat}\right] + br(t)N(t) - \mu N_s(t) - ar(t)N_s(t) - br(t)N_s(t) \quad (5)$$

where $N_T(t)=N(t)+N_s(t)+N_d(t)$ is the total number of cells at time *t*, which controls the effective proliferation rate.

A Euler method (Press et al 2007) was coded in to numerically solve the system of equations. At each time step, the solution is updated as:

$$N_i \rightarrow N_{i+1} = N_i + \frac{dN_i}{dt}\Delta t \quad (6)$$

The same holds for $N_s(t)$ and $N_d(t)$.

## 2.3. *The LQ-limit:* closed-form analytical solution and functional comparison with the LQ model

Rather than assuming an instantaneous dose rate (Dirac delta), we will study this limit differently, in a way that facilitates solving the system of equations. If dose delivery is fast compared to typical repair times, we can ignore the effect of sub-lethal damage repair. We also ignore proliferation and elimination of doomed cells. We suppose a constant dose rate *r(t)=d*. Under these approximations our system becomes:

$$\frac{dN}{dt} = -(a'+b')N \quad (7a)$$

$$\frac{dN_s}{dt} = b'N - (a'+b')N_s \quad (7b)$$

$$\frac{dN_d}{dt} = a'N + (a'+b')N_s \quad (7c)$$

where *a'=a\*d* and *b'=b\*d*.

The solution to the first equation is,

$$N(t) = N_0 \exp\left(-(a'+b')t\right) \quad (8)$$

And using that the total dose *D* is given by *D=d\*t*:

$$N(D) = N_0 \exp\left(-(a+b)D\right) \quad (9)$$

Inserting this equation in the differential equation for $N_s$ and solving it, we get the following solution:

$$N_s(t) = N_0 \frac{b't}{\exp(a't+b't)} \rightarrow N_s(D) = N_0 \frac{bD}{\exp((a+b)D)} \qquad (10)$$

The surviving fraction of cells will be $(N+N_s)/N_0$, as doomed cells will eventually die, and damaged cells will eventually recover if they do not suffer further damage. Therefore:

$$SF(D) = \exp(-(a+b)D)[1+bD] \qquad (11)$$

This expression is similar to the closed-form solution of Curtis (1985) model, even though some constants are different due to the use of different repair rates. This expression is formally different that the LQ model. However, a Taylor series analysis of this equation can help to relate both models. Expanding the exponent around $D=0$ (low dose limit), and comparing to the LQ equation, we find:

$$-\alpha D - \beta D^2 = -aD - bD + \left[bD - \frac{b^2 D^2}{2} + \frac{b^3 D^3}{6} - \cdots\right] \qquad (12)$$

Both models are equivalent up to order 2 with $\alpha=a$ and $\beta=b^2/2$, but beyond this limit, the functional dependence of both expressions is different, which is again similar to the low-dose comparison of Curtis' and LQ models.

**2.4. Fit to experimental data: surviving fractions and comparison to the LQ model**

We have shown in the previous section that a closed-form expression solution to our system under assumptions similar to those of the LQ model, is very similar to the LQ model in the low dose regime, but they functionally differ beyond such regime. However, we want to fit both models to experimental data to check if equation (11) can reproduce experimental surviving fractions.

Unkel *et al* (2016) reported surviving fractions for 9 cells lines presenting different radiosensitivities and very different $\alpha/\beta$ ratios when fitted to the LQ model. We have extracted reported surviving fractions, and fitted our model and the LQ model to the experimental data, in order to intercompare results. As those experimental results correspond to fast *dose delivery,* we have used the limit of our model explored in section 2.3 to fit the data. Best fitting parameters ($\alpha$ and $\beta$ in the LQ-model and $a$ and $b$ in our model) have been obtained by minimizing the following objective function:

$$F = \sum_i \left[\log(SF_{exp,i}) - \log(SF_{th,i})\right]^2 \qquad (13)$$

where $\{SF_{exp,i}\}$ and $\{SF_{th,i}\}$ are the experimental and theoretical surviving fractions. Simulations have been performed considering a constant dose rate of 6 Gy/min (0.1 Gy/s).

**2.5. Fit to experimental data: dynamic evolution of tumor volumes during metabolic radiotherapy**

We have used the model presented in this work to fit experimental data of evolution of tumor volumes treated with metabolic radiotherapy reported by Baiu et al (2019). In that publication, the authors reported the evolution of xenograft tumors of 4 different cell lines (Rh30, alveolar rhabdomyosarcoma; TC-71, Ewing sarcoma; CHLA-20 and NB-1691, neuroblastoma) treated with a single inoculation of $^{131}$I-CLR1404 or an excipient (control group). The authors also investigated the uptake and pharmacokinetics of the radiopharmaceutical by the different tumors, and were able to calculate radiation doses delivered to the tumor by MBq of injected activity (2.40, 1.21, 1.10 and 0.92 Gy/MBq for Rh30, NB-1691, CHLA-20 and TC-71, respectively).

We used these values, as well as the mean injected activities per body weight (115 MBq/kg) and an average body weight of 30 g, to obtain the average doses for each experiment. Average cumulative doses were 8.28, 4.17, 3.80 and 3.16 Gy for Rh30, NB-1691, CHLA-20 and TC-71, respectively. Dose rate curves were reconstructed by using pharmacokinetic curves presented in that article (uptake and biological clearance) and the physical decay of $^{131}$I. Dose rate curves are mainly conditioned by the physical decay of iodine, as the biological clearance of the radiopharmaceutical is slower than the physical decay (experimental data in Baiu et al (2019) shows little biological clearance up to 7 days post-administration). In order to reconstruct dose rate curves, we assumed that dose rate was proportional to activity concentration in the tumor.

Fitting was performed by minimizing the weighted sum of squared differences between experimental data and model data. A simulated annealing algorithm was implemented for this purpose (Kirkpatrick et al 1983). Tumor volumes were supposed to be proportional to the number of cells, with a cell density of cells of $10^8$ cm$^{-3}$ (Del Monte 2009). Some parameters were set to fix values, in particular $\gamma$=0.069 days$^{-1}$ (doomed cells have a half-life of 10 days), and $\mu$=0.006 min$^{-1}$ (repair rate with a half-life of 120 minutes, within the range of values reported in Steel et al (1987)), and only radiosensitivity parameters, *a* and *b*, the proliferation rate, $\lambda$, and the proliferation saturation $1/N_{sat}$, were allowed to change. Only undamaged cells were considered proliferative. While this set of simplifications certainly constitutes an approximation, as different cell lines may, for example, have different repair capabilities, it serves to avoid over-fitting of the data and can be enough to provide a qualitative fit to the experimental data.

**2.6. General application to molecular therapy: evaluation of tumor control probabilities**

We have used our model to illustrate the dynamics of populations of cells in a tumor irradiated by dose rate profiles similar to those received in molecular radiotherapy. In such studies, it is interesting to study tumor control probability (TCP), which assuming the clonogen hypothesis, is given by the probability to eliminate every single tumor stem cell. In this situation, it seems more appropriate to work with integer numbers of cells rather than with surviving fraction: the surviving fraction will never reach zero, and due to proliferation will eventually increase when the dose rate decreases, leading to fictitious very low TCP value. This can be solved by converting the deterministic system of equations in a stochastic system of equations (Badry and Leder 2016, Hanin 2001). In this work, this has been simply done by making the substitution:

$$\Delta N = \frac{dN}{dt} \Delta t \rightarrow \Delta N' = poissrand\left(\Delta N\right) \tag{14}$$

The variation of cells in each time step, which is calculated as *(dN/dt)Δt* is now converted to a Poisson random number (poissrand), and this (integer) random number is added to the population of cells (also integers). This is applied to each population of cells, *N*, $N_d$, and $N_s$.

We have qualitatively investigated the effect of repair and proliferation rates in the dynamics of populations of cells and tumor control. We have used dose rate profiles similar to those of $^{131}$I therapy, with a fast dose rate increase (tumor uptake), and slow decay (half-life ~8 days). For these simulations, we have ignored the contribution of biological decay to dose rate curves.

We have investigated proliferation rates from moderate to fast for undamaged cells, $\lambda$=0.069, 0.139, and 0.347 days$^{-1}$ (which correspond to potential doubling times of 10, 5, and 2 days), values that are within the reported ranges for experimental tumors (Kim and Tannock 2005, Baiu et al 2019). We have also set: $1/N_{sat}$=0 (proliferation is purely exponential); no proliferation for $N_d$ and $N_s$; an elimination rate of doomed cells $\gamma$=0.069 days$^{-1}$ (doomed cells have a half-life of 10 days); repair rates of damaged cells, $\mu$=0.011 min$^{-1}$ (60 minutes half-life), and $\mu$=0.006 min$^{-1}$ (120 minutes half-life). These repair rates are in the moderate to slow range for human tumor cells (Steel et al 1987).

We have also simulated TCP curves, with the methodology presented above. In order to compute TCP values for a given configuration (cumulative dose and cell response parameters), we performed 200 stochastic simulations, and the TCP is defined as the fraction of simulations that lead to control ($N$=0).

### 2.7. Implementation

The model and the optimization algorithm were implemented in Matlab (The Mathworks, Natwick, MA). Data and code are available from the Mendeley database (Neira et al 2020). Regarding the particular implementation of the Euler method (Eq. 6), the time step Δt was generally set to 1 minute, a value well below the minimum characteristic time of the biological effects included in the model. However, for the fitting to experimental data reported in Baiu et al (2019) (section 2.5) and the calculation of stochastic tumor control probabilities (section 2.6), that require a very large number of simulations, Δt=10 minutes was used instead. Such time steps are well below the typical times of the fastest process in our model (60 minutes half-life for sub-lethal damage repair) and do not affect the precision of the solution of our system of equations. However, lower Δt values should be used if investigating very fast recovery rates.

## 3. Results and discussion

### 3.1. Fits to experimental data: surviving fractions and comparison to the LQ model

In Figure 1 we present best fits to surviving fractions for 9 cells lines for the LQ model and equation (11). The figure is complemented by Table 1, where we show best-fitting parameters and the value of the optimized cost function.

In general, results obtained with equation (11) match very well the LQ model: 3 fits are identical, in 4 fits equation (11) provides a slightly better fit, and in 2 cases the LQ provides a better fit. Of the latter two cases, one of them (Suit-2 007 cell line) is especially interesting. For this cell line, the α/β ratio is particularly low (~1 Gy), and the LQ fit is significantly better than the fit obtained with equation (11). We have numerically investigated this issue, and we have found that equation (11) has problems to reproduce data with a low $\alpha/\beta$ ratio. This is a limitation of the model, even though most cancer cell lines tend to have large $\alpha/\beta$ ratios, and the model is intended to be used for tumors.

In the fits, the parameter *b* dominates cell death (i.e. most cell death is due to the accumulation of sub-lethal damage rather than due to direct lethal events). In fact, in several fits *a* tends to zero. Certainly, this constitutes a very important difference with the LQ model, as in the *infinitesimally low dose rate delivery* the LQ would tend to SF=exp(-α/D) (all sub-lethal damage is repaired), while the solution of our model in that limit with *a*=0 is SF=1.

### 3.2. Fits to experimental data: dynamic evolution of tumor volumes during metabolic radiotherapy

In Figure 2 we present best fits to the evolution of volumes of xenograft tumors of different cell lines after metabolic radiotherapy with $^{131}$I-CLR1404. Results are shown for the control and treated groups of mice. The model can qualitatively fit the evolution of tumor volumes, both in the control group and the treated groups.

### 3.3. General application to molecular therapy: evaluation of tumor control probabilities

We studied the evolution of populations of cells in a tumor, initially consisting of $10^6$ cells, when irradiated continuously, with a dose rate profile typical of $^{131}$I therapy (half-life~8 days), and a total cumulative dose of 110 Gy. We have investigated two different radiosensitivities: $a$=0 Gy$^{-1}$, $b$=0.664 Gy$^{-1}$, which correspond to Patu-8988T cells with effective values $α$=0.235 Gy$^{-1}$, $β$=0.025 Gy$^{-2}$ (see Table 1), and; $a$=0.473 Gy$^{-1}$, $b$=0.206 Gy$^{-1}$, which correspond to Dan-G cells with effective values $α$=0.532 Gy$^{-1}$, $β$=0.005 Gy$^{-2}$ (see Table 1).

In Figure 3 we show results for Patu-8988T cells. Initially, the high dose rate at the beginning of the treatment greatly affects the population of healthy cells. However, combinations of fast repair and/or fast proliferation rates can eventually balance the diminishing effect of radiation (due to radioisotope decay), the tumor can recover and there is no control. While tumor bulk is still observed for a long time due to the slow elimination of doomed cells, local control is assured as there are no viable cells in it. For slow repair/proliferation rates this does not happen, and the tumor is effectively controlled.

In Figure 4 we present results for Dan-G cells. The particularity of this cell line is that the amount of repairable sub-lethal damage is low (low *β* and *b* values). Therefore the tumor is insensitive to the protraction effect of prolonged dose delivery, and tumor control can be achieved for every investigated combination of repair/proliferation rates.

This is further investigated in Figure 5, where we present simulated TCP vs cumulative dose curves for the cell lines investigated above: Patu-8988T tumors are more radio-resistant to prolonged dose delivery, and present $D_{50}$ (cumulative dose necessary to achieve 50% control) values around 140 Gy, while Dan-G tumors are insensitive to the protraction effect of prolonged dose delivery, and present $D_{50}$ ~ 40 Gy.

## 4. Conclusions

We present a simple kinetic model of radiation damage to populations of cells, including populations of viable tumor cells, sub-lethally damaged tumor cells (which can repair), and doomed (non-repairable) cells. The model includes proliferation, sub-lethal and lethal (either directly or by a combination of sub-lethal events) damage, repair, and elimination of doomed cells. We consider a logistic-like proliferation term, but other types of proliferation can be easily accommodated in the model. Repair of sub-lethally damaged cells is exponential, but other functional forms can be easily included in the model, like e.g. Michaelis-Menten repair kinetics (Murray 2003), or bi-exponential kinetics (Dale 2018).

In this piece of work, we have consistently considered that only viable cells can proliferate and that the radiosensitivity of viable and sub-lethally damaged cells is the same (same *a* and *b* parameters in our model). This was done for the sake of simplicity, to avoid having a large number of free parameters that can cause overfitting of the experimental data. Certainly, this has to be considered an approximation, as radiosensitivity of sub-lethally damaged cells may be different, and doomed cells may still keep some proliferative capacity (and so do sub-lethally damaged cells, but because repair is much faster than proliferation, the population of sub-lethally damaged cells is bound to be low and the contribution of a proliferation term for these cells will be minimal).

The main advantage of the model that we present in this work is that it incorporates in an easy way different cross-interacting compartments of cells forming a tumor. Therefore, the model may be especially interesting for studying the dynamics of treated tumors, including tumor control probabilities, tumor shrinking, re-oxygenation … For example, important properties of tumor dynamics, such as shrinking and re-oxygenation, could be linked to fractions of living cells (healthy, damaged, and doomed), and radiosensitivity adjusted accordingly during treatment according to re-oxygenation (Wouters and Brown 1997). The same holds for accelerated proliferation, which may be triggered when the number of cells in a tumor gets below a given threshold (Pedicini 2013). While this issue has not been explored in this note, such implementation would be direct in this model.

In the *fast dose delivery limit* (dose delivery time is much shorter than typical repair and proliferation times), the model can be analytically solved, obtaining a simple closed-form expression that is very similar to the solution to Curtis (1986) model. We have investigated this solution, finding that it is equivalent to the LQ formulation in the low-dose range, but in general, they are functionally different. However, fitting of several surviving curves with both models shows that they can produce similar fits. There is one limitation with our model, though, it cannot fit adequately surviving fractions of cells with a low *α/β* ratio. This intrinsic limitation, while theoretically concerning, may not be a serious practical limitation as most tumor cells present large *α/β* ratios. Interestingly, in several fits *a* tends to zero, which implies that cell death is mostly due to accumulation of sub-lethal damage.

Kinetic models like the one presented in this work may be of special interest when modeling response to molecular radiotherapy. In this type of technique, an activity of a given radionuclide is taken up by the tumor, where it decays creating a continuous dose rate irradiation which depends on the physical decay kinetics of the radionuclide and the biological kinetics of uptake/elimination. We have shown that the model can qualitatively fit the evolution of tumor volumes after molecular radiotherapy, by fitting experimental data of different cell-line xenograft tumors in mice treated with $^{131}$I-CLR1404. We have also illustrated the application of our model to this particular problem, simulating the response of a tumor to a $^{131}$I-like treatment, and showing how tumor response parameters can affect the evolution of the tumor and the probability of tumor control.

# Acknowledgments

This project was supported by Instituto de Salud Carlos III (ISCIII) through grants CPII17/00028,DTS17/00123 and PI17/01428 (FEDER co-funding). This project has received funding from the EuropeanUnion's Horizon 2020 research and innovation programme under the Marie Skłodowska-Curie grantagreement No 839135.

# Declarations

Ethics approval and consent to participate: Not applicable.

Availability of data and materials: Code and data are shared in the Mendeley database.

Competing interests: The authors declare that they have no competing interests.

| Cell line | LQ model | | | Equation (11) | | |
|---|---|---|---|---|---|---|
| | $\alpha$ [Gy$^{-1}$] | $\beta$ [Gy$^{-2}$] | cost | $a$ [Gy$^{-1}$] | $b$ [Gy$^{-1}$] | cost |
| L3.6pl | 0.428 | 0.025 | 2.17 | 0.007 | 0.866 | 1.61 |
| MiaPaca-2 | 0.241 | 0.037 | 2.40 | 0.000 | 0.777 | 2.99 |
| Panc-1 | 0.332 | 0.020 | 1.24 | 0.099 | 0.606 | 1.13 |
| Patu-8988T | 0.235 | 0.025 | 1.74 | 0.000 | 0.664 | 1.51 |
| Suit-2 007 | 0.066 | 0.058 | 1.32 | 0.000 | 0.789 | 3.62 |
| Capan-2 | 0.593 | 0.031 | 6.29 | 0.001 | 1.091 | 5.08 |
| Dan-G | 0.532 | 0.005 | 4.59 | 0.473 | 0.206 | 4.55 |
| FamPac | 0.813 | 0.003 | 2.99 | 0.775 | 0.104 | 3.00 |
| HPDE | 0.259 | 0.034 | 8.11 | 0.000 | 0.774 | 8.17 |

Table 1: Best fitting parameters and optimized value of the cost function of fits to experimental surviving fractions of nine cell lines (taken from Unkel *et al* 2016) with the LQ model and Equation (11).

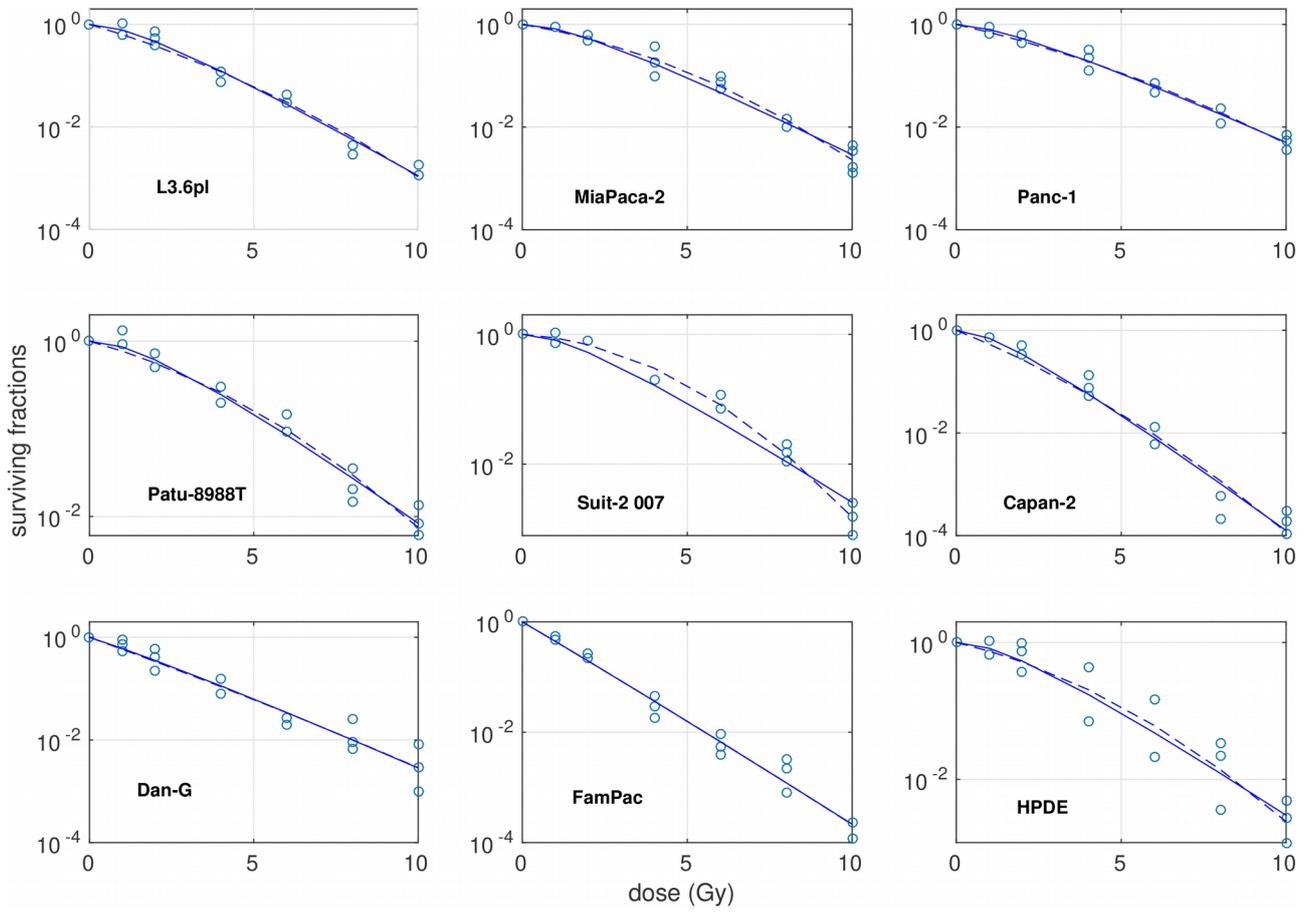

Figure 1: Experimental surviving fractions (circles) versus dose for nine different cell lines, and best fits obtained with the LQ model (dashed lines) and equation (11) (solid line). Best fitting parameters are reported in Table 1.

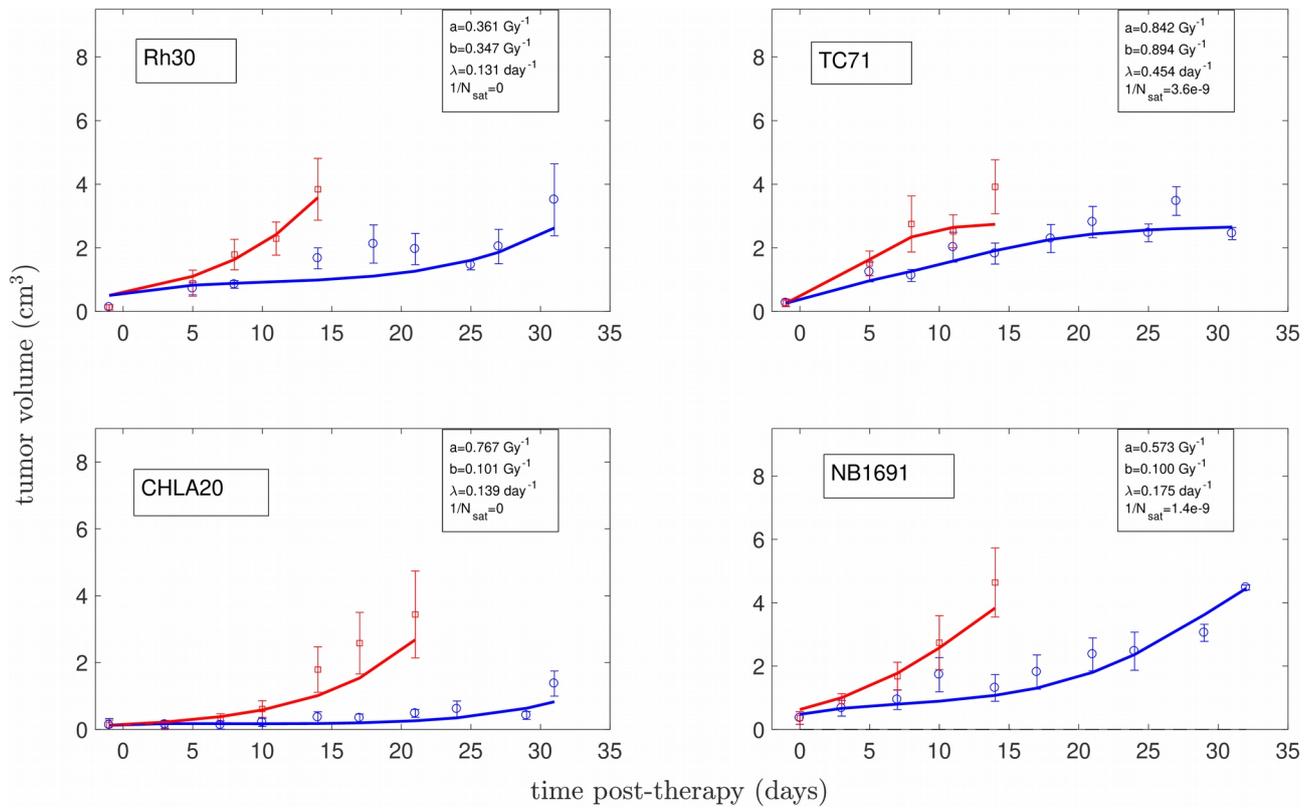

Figure 2: Experimental evolution of xenograft tumor volumes of four different cell lines (Rh30, TC71, CHLA20, NB1691) in mice after the administration of radio-iodine $^{131}$I-CLR1404 therapy (circles) or excipient (squares), and best fits obtained with the model presented in this work (solid lines). Best-fitting parameters are shown in the figure. Parameters that are not included in the figure were fixed to the following values: $\gamma=0.069$ days$^{-1}$ (doomed cells have a half-life of 10 days); $\mu=0.006$ min$^{-1}$ (repair rate with half-life of 120 minutes).

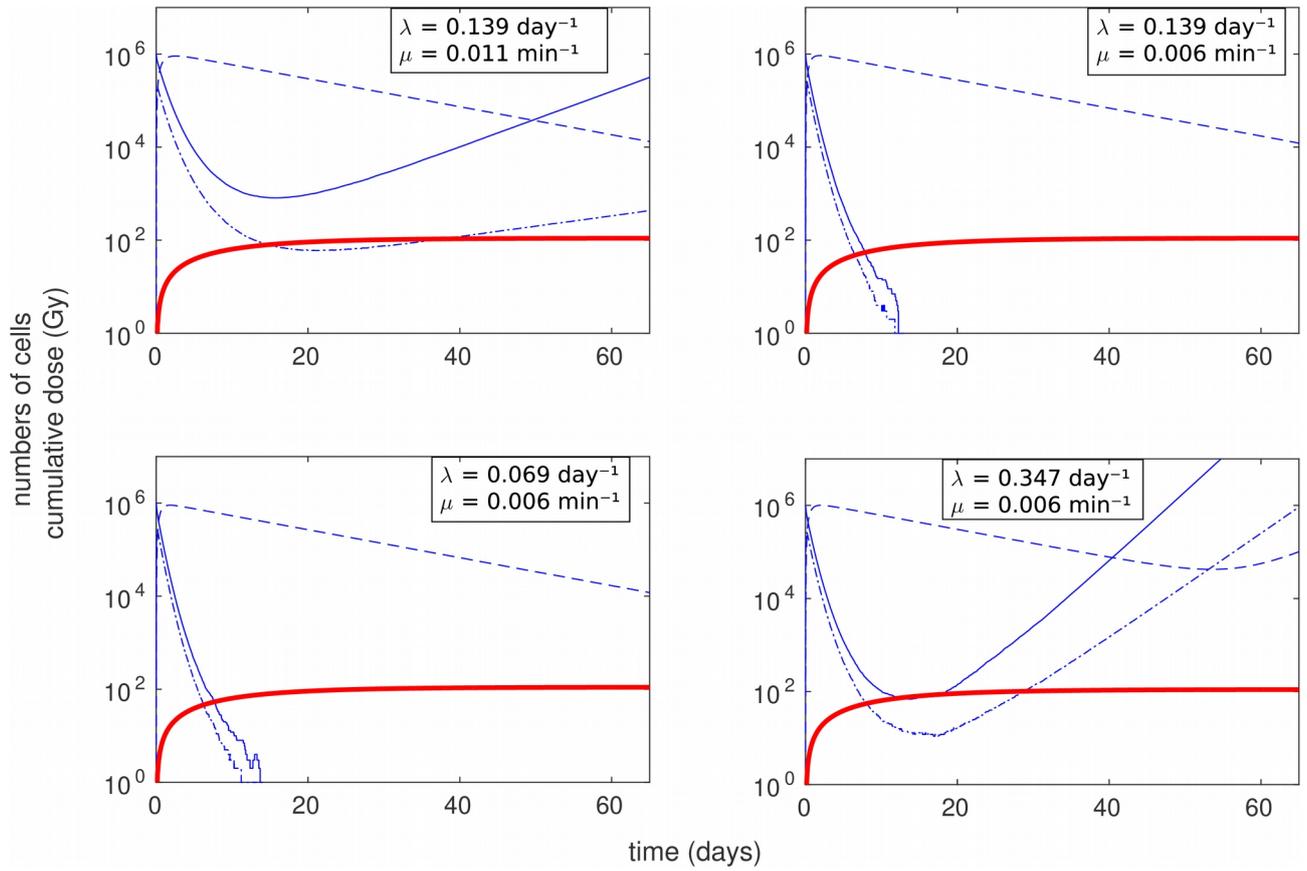

Figure 3: Evolution of the number of cells (viable, $N$, thin solid lines; sub-lethally damaged, $N_s$, dashed lines; doomed, $N_d$, dash-dot lines), with time in a tumor irradiated by a $^{131}$I-like dose rate profile. The thick solid line shows the absorbed dose versus time (total cumulative dose~110 Gy). Radiosensitivites correspond to Patu-8988T cells in Table 1 (a=0.0 Gy$^{-1}$, b=0.664 Gy$^{-1}$; $\alpha$=0.235 Gy$^{-1}$, $\beta$=0.025 Gy$^{-2}$). Doomed cells have a half-life of 10 days ($\gamma$=0.069 days$^{-1}$). Different proliferation and sub-lethal damage repair rates are investigated (indicated within each subpanel). Cells that proliferate fast or recover fast can avoid tumor control (top-left and bottom-right panels).

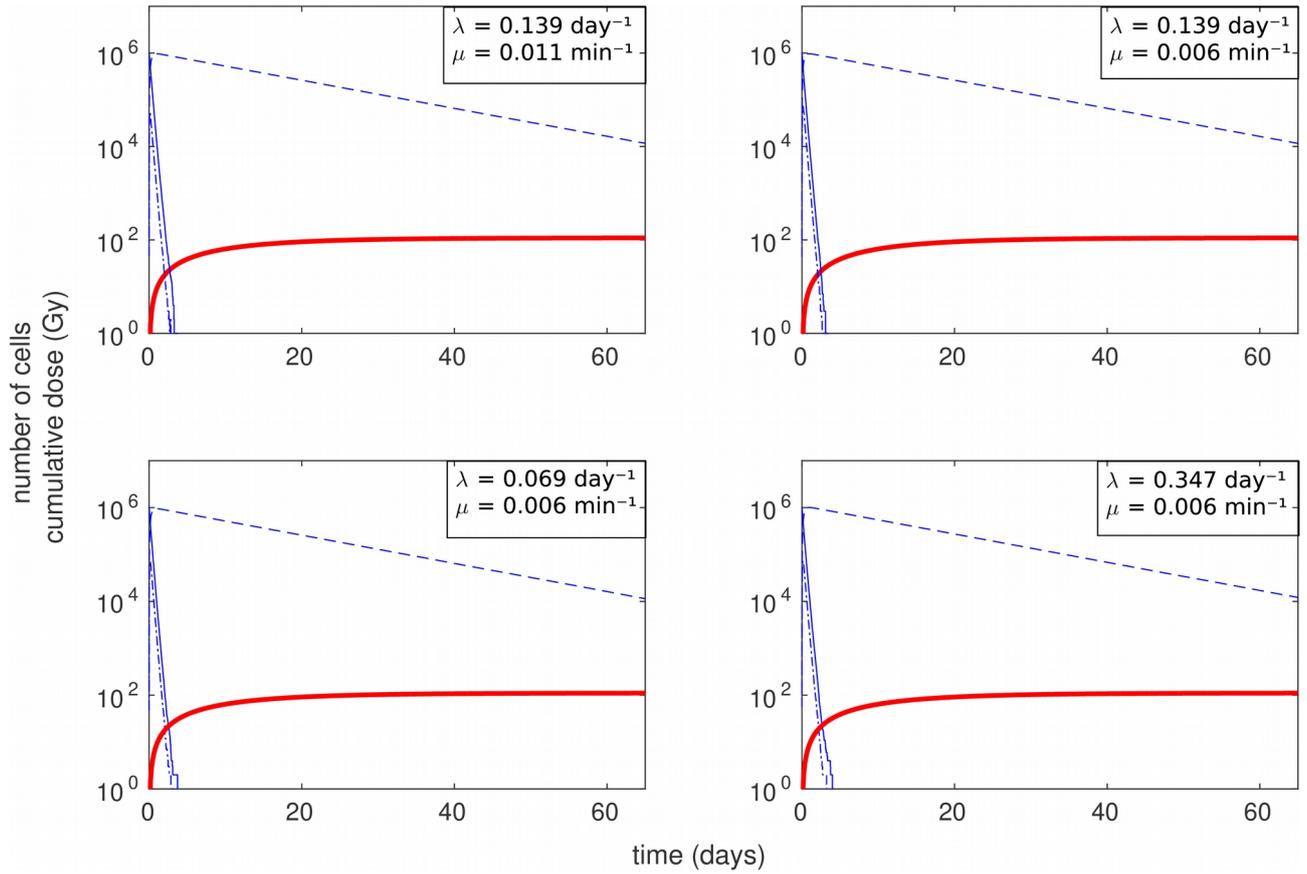

Figure 4: Evolution of the number of cells (viable, $N$, thin solid line; sub-lethally damaged, $N_s$, dashed lines; doomed, $N_d$, dash-dot lines), with time in a tumor irradiated by a $^{131}$I-like dose rate profile. The thick solid line shows the absorbed dose versus time (total cumulative dose ~110 Gy). Radiosensitivites correspond to Dan-G cells in Table 1 (a=0.473 Gy$^{-1}$, b=0.206 Gy$^{-1}$; $\alpha$=0.532 Gy$^{-1}$, $\beta$=0.005 Gy$^{-2}$). Doomed cells have a half-life of 10 days ($\gamma$=0.069 days$^{-1}$). Different proliferation and sub-lethal damage repair rates are investigated (indicated within each subpanel). For this cell line, the amount of repairable damage is low, and tumor control is achieved with this cumulative dose for every investigated combination of proliferation/sub-lethal damage repair rates.

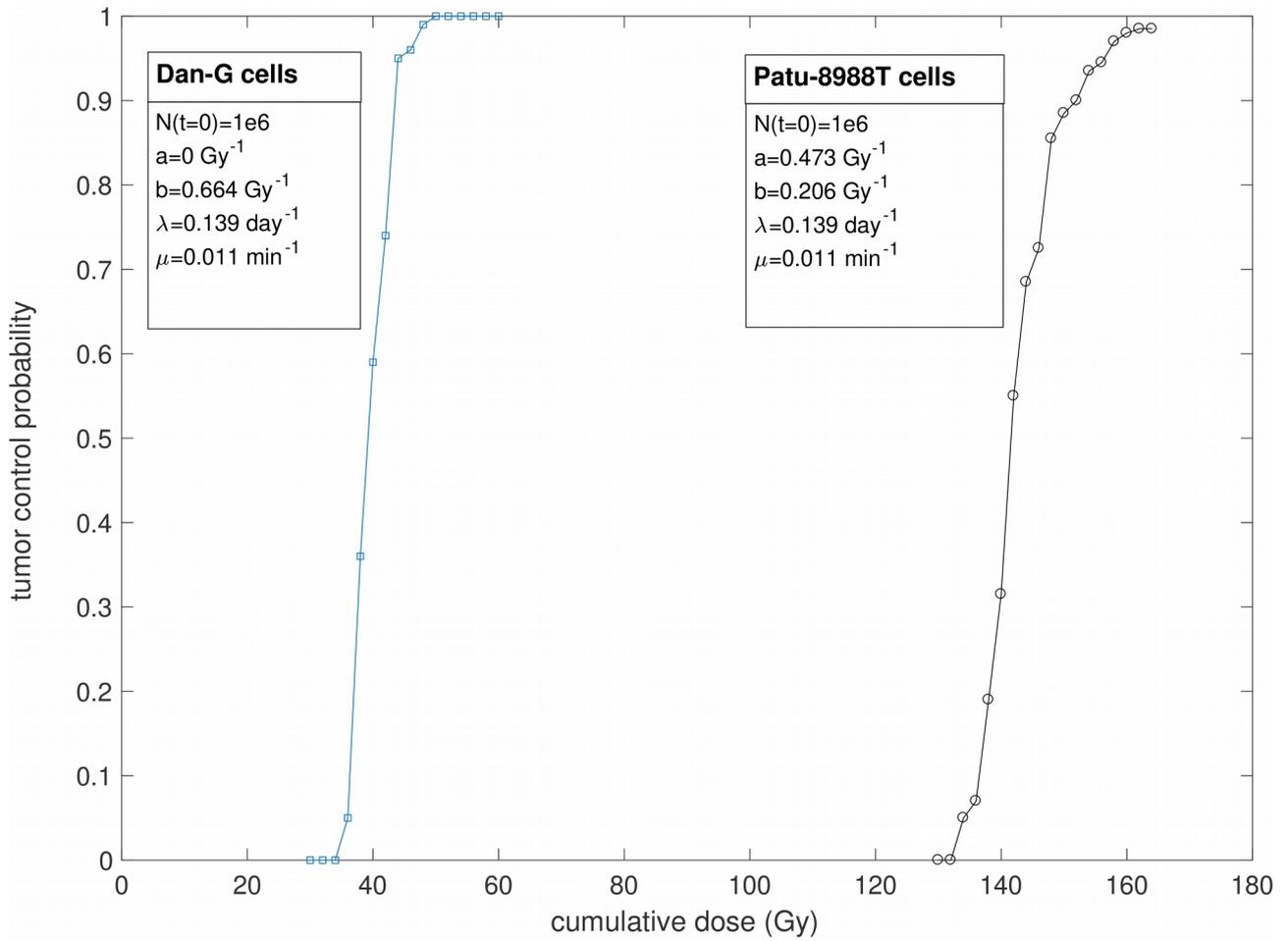

Figure 5: Simulated TCP vs cumulative dose (delivered with a $^{131}$I-like dose rate profile) for Dan-G and Patu-8988T. Radiosensitivity, proliferation and repair parameters used in the simulations are provided within the figure. Each TCP point was calculated by performing 200 stochastic simulations.